# Less is more – enhancement of second-harmonic generation from metasurfaces by reduced nanoparticle density

*Robert Czaplicki[†,‡]\*, Antti Kiviniemi[†], Mikko J. Huttunen[†], Xiaorun Zang[†], Timo Stolt[†], Ismo Vartiainen[§], Jérémy Butet[¥], Markku Kuittinen[§], Olivier J. F. Martin[¥], and Martti Kauranen[†]\**

[†] Laboratory of Photonics, Tampere University of Technology, P. O. Box 692, FI-33101 Tampere, Finland

[‡] Institute of Physics, Faculty of Physics, Astronomy, and Informatics, Nicolaus Copernicus University, Grudziądzka 5/7, 87-100 Toruń, Poland

[§] Institute of Photonics, University of Eastern Finland, P. O. Box 111, FI-80101 Joensuu, Finland

[¥] Nanophotonics and Metrology Laboratory (NAM), Swiss Federal Institute of Technology, Lausanne (EPFL), 1015 Lausanne, Switzerland

ABSTRACT: We investigate optical second-harmonic generation (SHG) from metasurfaces where noncentrosymmetric V-shaped gold nanoparticles are ordered into regular array configurations. In contrast to expectations, a substantial enhancement of the SHG signal is observed when the number density of the particles in the array is reduced. More specifically, by halving the number density, we obtain over five-fold enhancement in SHG intensity. This striking result is attributed to favorable interparticle interactions mediated by the lattice, where surface-lattice resonances lead to spectral narrowing of the plasmon resonances. Importantly, however, the results cannot be explained by the improved quality of the plasmon resonance alone. Instead, the lattice interactions also lead to further enhancement of the local fields at the particles. The experimental observations agree very well with results obtained from numerical simulations including lattice interactions.

Surface plasmons are collective oscillations of conduction electrons, which determine the optical responses of metals. When surface plasmons are excited in metal nanoparticles, they become localized and can exhibit resonant behavior. These resonances, known as localized surface plasmon resonances (LSPRs), can strongly enhance the local optical fields near the particles [1]. Since the local fields rule the light-matter interactions at the nanoscale, their enhancement has facilitated the development of numerous applications based on metal nanoparticles, ranging from optical antennas, perfect lenses, light trapping structures in solar cells, surface enhanced spectroscopies to sensing [2-9]. Importantly, the strengths of light-matter interactions can be further boosted by arranging nanoparticles into periodic lattices, also known as metasurfaces. In such systems, the nanoparticles are coupled to each other either through near-field interactions [10] or through their scattered fields. In the latter case, the samples can exhibit collective responses known as surface-lattice resonances (SLRs) with modified spectral features [11-16].

Nonlinear optics is crucial in many photonic applications ranging from entangled-photon generation [17] to frequency combs [18]. However, the intrinsic material nonlinearities are often very weak, making it challenging to realize efficient nonlinear photonic devices with small footprints and reduced power requirements. The possibility to enhance local fields by utilizing LSPRs and SLRs of metasurfaces can thus be especially important for nonlinear optics, where the light-matter interactions scale with higher powers of the fields. For example, second-harmonic generation (SHG) is a nonlinear optical process where two input fields oscillating at a fundamental frequency are combined into an output field oscillating at the doubled frequency. The efficiency of the SHG process scales with the fourth (second) power of the input field amplitude (intensity) and is thus very sensitive to changes in the local fields [19].

The nonlinear properties of metal nanoparticles have been investigated for individual particles [20-25] and metasurfaces [10,26-35]. In the latter case, one can envision two strategies to enhance the overall nonlinear response. By treating each nanoparticle as an elementary source of coherent nonlinear radiation, one expects the nonlinear response to scale with the square of the particle number density $N$. On the other hand, by operating close to the plasmon resonance of the nanoparticles, the response is expected to depend also on the quality of the resonance [35,36]. An attempt has been made to enhance the overall response by increasing the number of nanoparticles on a metasurface. Unfortunately, however, this compromised the quality of the resonance through interparticle interactions [10], limiting the achievable nonlinearity through this approach.

It therefore appears that any further enhancements in nonlinear responses need to be based on improving the quality of the resonances, i.e., on designing metasurfaces with very narrow linewidths. However, the possibilities to reduce the linewidths associated with LSPRs to a greater degree are seemingly limited, because metals are intrinsically lossy materials [2,3,37]. Despite this common belief, this is not a fundamental limitation, since utilization of SLRs can result in spectral features with remarkably narrow linewidths [12,38,39]. In spite of these opportunities, only a few reports exist where SLRs have been used to enhance nonlinear responses of metasurfaces [40-42]. However, recent theoretical predictions based on nonlinear discrete-dipole approximation (DDA) suggest that the nonlinear responses of metasurfaces can be enhanced by several orders of magnitude in the presence of SLRs [39,42].

In this Letter, we provide a striking example how SLRs can boost nonlinear responses. We demonstrate that a decrease in the particle number density of a metasurface can significantly enhance its nonlinear response. In particular, we show that the SHG efficiency of a metasurface is enhanced over five-fold by halving the number of nanoparticles of the metasurface compared to a reference array. Although both arrays consist of identical V-shaped gold nanoparticles and exhibit similar linear responses, the arrangement of the elements in the modified metasurface results in favorable interparticle interactions and in the enhancement of SHG. The experimental results are validated by finding a good agreement with numerical simulations based on nonlinear DDA approach [42] and further confirmed by simulations based on surface integral equations [34] and finite element method [43].

We fabricated two-dimensional arrays of gold nanoparticles by using electron-beam lithography and lift-off techniques. The 20-nm thick nanoparticles were separated from the fused silica ($SiO_2$) substrate by a 3-nm thick adhesion layer of chromium and were covered from the top by a 20-nm thick protective layer of $SiO_2$. Note, that such a thin dielectric layer on top of the nanoparticle array also facilitates coupling of LSPRs to diffractive orders of the array, thus favoring formation of SLRs [14]. The nanoparticles were designed to be V-shaped nanoantennas consisting of two equal arms (with length $l = 275$ nm, and width $w = 100$ nm) oriented to each other to form an angle of 90° [see Fig. 1(a)]. The particles were arranged into square lattices with periods $p_x = p_y = 500$ nm. In the reference sample (V1), all the lattice points of the array were filled with particles, resulting in a square unit cell with an area of 500 × 500 nm², occupied thus by one particle [see dotted-line square in Fig. 1(a)]. The second sample (V2) was designed by removing the particles from every second lattice point (both in $x$- and $y$-directions), resulting in halved particle number density $N$ compared to the reference sample V1, which was thus expected to result in four-fold decrease in the detected SHG signal. In other words, the remaining particles formed a new lattice rotated by 45° from the original one, with new periods of ~707 nm in the directions $-x+y$ and $x+y$ [see dotted-line square in Fig. 1(b)]. The V-shaped particles support LSPRs excited most efficiently by two orthogonally oriented input polarizations: along the $x$- and $y$-directions [44], as determined by the symmetry of the particles.

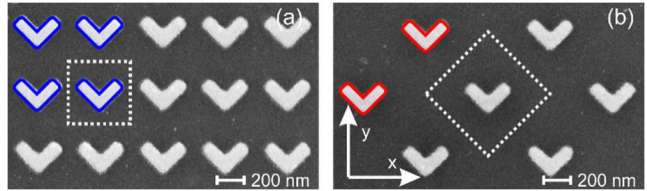

Figure 1. The scanning electron microscopy images of the reference sample V1 (a) and the modified sample V2 (b). The images show a portion of the sample area with 5×3 lattice points of the square array with 500 nm period. Blue (a) and red (b) outlines of nanoparticles highlight the number of particles in a 1×1 µm² area. Dotted-line squares indicate the unit cells of the arrays. The coordinate system used for all samples is shown as an inset in (b).

We first characterized the samples by linear extinction measurements. These measurements were performed by illuminating the samples with a collimated beam (diameter ~1 mm) from a halogen lamp and using linearly-polarized light along *x*- and *y*-directions. The spectral position of the *y*-polarized resonance is in the spectral window of 1000 – 1300 nm [Fig. 2(a)], which matches closely to the fundamental wavelength used in the SHG measurements. These results also agree very well with the numerical simulations (see Fig. S1 in Supporting Information). Note that the *y*-polarized spectrum of the modified sample V2 is reduced in strength, due to reduced particle density *N*. However, the reduction in extinction is less than the reduction in number density. In addition, the resonance of the modified sample V2 is red-shifted and its linewidth is narrower than that of the reference V1. All these effects provide evidence of the excitation of SLRs in sample V2. In particular, the arrow in the Fig. 2(a) shows the position of the Rayleigh anomaly (RA) for sample V2 that is related to the diffraction orders (±1,0), (0,±1) of the array. Other possible RAs, outside the spectral range of our interest (1000 – 1300 nm), are not shown. The described RA is responsible for the formation of SLR in the case of sample V2 and is also the origin of the redshift of the *y*-polarized resonance with respect to the resonance of sample V1 [40].

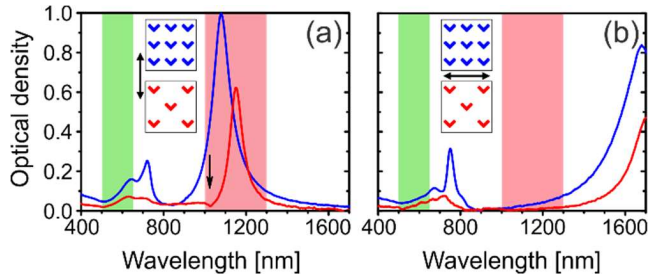

Figure 2. Extinction spectra for V1 (blue) and V2 (red) samples for (a) *y*- and (b) *x*-polarized excitation (the direction of polarization is indicated by the double-side arrows next to the sample designs). The colored red (green) areas highlight the fundamental (SHG) wavelength range.

The nonlinear properties of the samples were studied by measuring their second-harmonic (SH) responses. A normally incident fundamental beam was provided by a pulsed laser (Chameleon

Vision II, Ti:Sapphire, 80 MHz, pump wavelength: 770 nm) combined with optical parametric oscillator (OPO; Chameleon Compact, 1000 – 1300 nm, pulse length: 200 fs) and SH light was detected in transmission (Fig. 3). The power of the fundamental beam from the OPO was controlled by a motorized achromatic half-wave plate (HWP) and a polarizer. A set of lenses and an aperture (diameter 25 μm) were used to clean and expand the beam before entering the polarization-control part of the setup. We used an achromatic lens of 150 mm focal length to weakly focus the beam on the sample arrays. This ensured that the size of the excitation beam was relatively small (diameter of the beam waist was ~100 μm) while still being close to a plane wave. A high-quality polarizer and an achromatic HWP were used to control the input polarization, while a film polarizer after the sample was used to select the polarization of the emitted SHG light. A 900 nm long-pass (700 nm short-pass) filter was used to pass (block) the fundamental beam. A lens of focal length of 16 mm was used after the sample to efficiently collect the generated SH signal. Another achromatic lens of 150 mm focal length focused the SHG signal on the active area of a photomultiplier tube (PMT) module after being reflected by a dichroic mirror and passing through another short-pass filter (900 nm). The light transmitted through the dichroic mirror was used to image the sample plane with a CMOS camera and a camera lens (MVL50M23) for sample alignment.

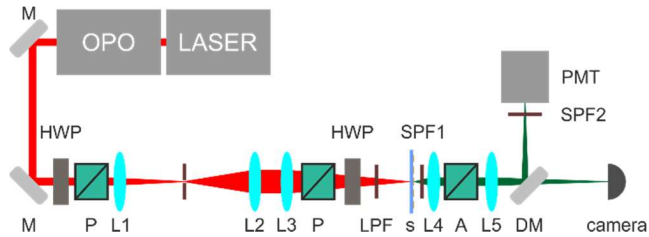

Figure 3. Schematic representation of the spectral SHG setup. M – mirrors, HWP – motorized half-wave plates, P – polarizers, L – lenses (L1, $f$=30 mm; L2, $f$=150 mm; L3, $f$=150 mm; L4, $f$=16 mm; L5, $f$=150 mm; L1, L2, L3, and L5 – achromatic), LPF – long-pass filter 900 nm, SPF1

– short-pass filter 700 nm, A – film polarizer (analyzer), DM – dichroic mirror, SPF2 – short-pass filter 900 nm, PMT – photomultiplier tube (PicoQuant PMA-C 192-M).

In order to gain a complete picture of the spectral SHG responses of the samples, we performed nonlinear experiments while scanning the wavelength of the fundamental (SHG) beam from 1000 to 1300 nm (from 500 to 650 nm) [see Fig. 4(a)]. The symmetry of the samples ($C_{1v}$) dictates that the non-vanishing second-order susceptibility tensor components are $\left(\chi_{yyy}^{(2)}, \chi_{yxx}^{(2)}, \chi_{xyx}^{(2)} = \chi_{xxy}^{(2)}\right)$. To simplify the analysis of our results, the following discussion is focused solely on the tensor component $\chi_{yyy}^{(2)}$, which is both allowed by symmetry and exhibits a resonance that we were able to experimentally access using the OPO (shaded red area in Fig. 2). At first glance, the SHG responses follow the intuitively expected behavior where the highest SHG efficiency occurs approximately at the wavelengths of the *y*-polarized resonances [1081 nm for V1 and 1151 nm for V2; compare Figs 2(a) and 4(a)]. The slight blueshift of the SHG maximum with respect to the maximum of the LSPRs agrees with previous observations [35]. Larger discrepancy from the expected behavior is observed in the case of sample V1, since the spectral SHG responses are broader and less intense than expected by looking at the respective linear response [see Fig. 2(a)]. A simple qualitative comparison of SHG responses shows clearly that the sample V2 exhibits a considerably stronger SHG response than the sample V1. More specifically, the response of V2 is enhanced by a factor of 5.4 comparing to the sample V1 at the respective maxima of the two samples. In addition, at the wavelength where both samples have equal optical densities (1135 nm), V2 demonstrates over eight-fold increase in the detected SHG signal [see grey dashed double-side arrow in Fig. 4(a)].

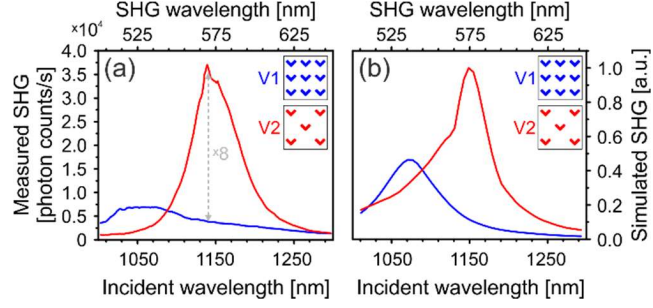

Figure 4. Measured (a) and simulated (b) wavelength-dependent emitted SHG signals for the metasurfaces V1 (blue) and V2 (red). Grey dashed double-side arrow shows the difference in the SHG at 1135 nm, where the optical densities of the arrays are equal. The simulations are performed with the DDA method.

The resonant spectral features of metasurfaces are well described by the Lorentz model [35,45]. Indeed, as a first approximation, close to the plasmon resonance at the fundamental wavelength, the SH responses of metasurfaces can be described by the relation:

$$\chi_{eff}^{(2)} \sim \frac{N}{(\Delta - i\gamma)^2}, \qquad (1)$$

where $N$ describes the number density of the nanoparticles, $\Delta$ is the detuning of the plasmon line center from the fundamental frequency and $\gamma$ is the linewidth of the resonance. Consequently, the detected SHG signal intensity is proportional to the square of Eq. (1). In this framework, the second-order susceptibility $\chi_{eff}^{(2)}$ on resonance ($\Delta = 0$) is proportional to $N$ and inversely proportional to the square of $\gamma$. Therefore, the resonant SHG response can be enhanced either by increasing the number density or by decreasing the linewidth of the resonance.

We first consider the results at the respective linecenters of the two samples (Table 1). Compared to the reference sample V1 with number density $N$ and linewidth $\gamma$, the modified sample V2 has number density $N/2$ and linewidth $0.7\gamma$. In consequence, the SHG signals from the two samples should be approximately equal (see Table 1). However, the second-harmonic intensities recorded for the two samples differ by a factor of five. This already suggests that the Lorentz model alone is not sufficient to explain this observation. Instead the local fields at the particles are also significantly modified by the modified interparticle interactions mediated by the SLRs.

Table 1. Parameters of the resonances and SHG responses [at linecenters (1081 nm – V1, 1151 nm – V2) and at 1135 nm] of samples V1 and V2. The scaling factors are calculated using Lorentz model [Eq. (1)] and the SHG intensities are obtained from the measurements (Fig. 4). The numbers are normalized to the respective ones for sample V1.

| Sample | $y$-polarized resonance (linecenter) [nm] | FWHM [nm] | Linewidth of the resonance | Number density $N$ | Normalized SHG scaling factor | | Normalized SHG intensity ($yyy$) | |
|---|---|---|---|---|---|---|---|---|
| | | | | | at linecenter | at 1135 nm | at linecenter | at 1135 nm |
| V1 | 1081 | 108 | $\gamma$ | 1 | 1 | 1 | 1 | 1 |
| V2 | 1151 | 76 | $0.7\gamma$ | 0.5 | 1.02 | 2.94 | 5.01 | 8.54 |

The spectral positions of resonances for both samples are different, therefore we performed similar analysis at the wavelength where both samples have equal optical densities (1135 nm). At this wavelength the analysis based on the Lorentz model [see Eq. (1)] must include the detuning of the resonances ($\Delta$ = 54 nm and $\Delta$ = 16 nm for sample V1 and V2, respectively). In consequence, due to the smaller detuning (0.3$\Delta$, comparing to sample V1) the SHG from the sample V2 should be enhanced by a factor of 3 (see Table 1). Instead, the SHG signal from sample V2 is stronger by a factor of 8.5. Such enhancement in the SHG response provides further support to the result that the lattice interactions, which are not accounted for in the Lorentz model, play a crucial role in the overall SHG responses and are responsible of the measured enhancement. It is important to underline, that the position of the RA at 1025 nm [Fig. 2(a)] results from the specific parameters of the sample V2 (period = 707 nm and refractive index of the substrate 1.45) favoring formation of a SLR, while no SLRs are formed in the case of sample V1 (period = 500 nm).

In order to verify the origin of these results, we performed simulations using an approach based on nonlinear discrete-dipole approximation (DDA) [39], which takes fully into account the lattice interactions, occurring both at the fundamental and the SHG wavelengths. In this approach, each particle in the array is treated as an electric dipole associated with a polarizability that is extracted from experimental data [Section 2 in Supporting information]. The simulated SHG signal [Fig. 4(b)] shows very good agreement with the experimental results [Fig. 4(a)]. The slight discrepancy arises from the fact that the nanoparticles of finite size are reduced to point dipoles and from experimental imperfections that are not fully accounted for. The excellent agreement between the experimental data and the nonlinear DDA simulations were further confirmed by performing simulations using other numerical tools, namely the finite element method (by COMSOL Multiphysics) and surface integral equation (SIE) method [Sections 3 and 4 in Supporting information]. These additional simulations were performed to provide further numerical support to the relatively novel DDA approach. All simulations show an enhancement of SHG when the lattice interaction is effective and the LSPR linewidth is reduced. This modification of the LSPR linewidth results in a three-fold enhancement of the fundamental intensity close to the nanoparticles, see Fig. S5. Since the fundamental near-field enhancement around each nanoparticle is stronger, the second-harmonic dipole induced in each nanoparticle is also stronger resulting in enhanced SH emission for sample V2, despite reduced nanoparticle density. In other words, a proper design of the lattice interactions and fundamental field enhancement is able to overcome a lower second-harmonic emitter density.

In conclusion, we have shown that second-harmonic generation from metasurfaces can be significantly enhanced by proper designs of the lattice interactions leading to strong light-matter interactions in the arrays through enhanced local fields. We have verified this principle by investigating two metasurfaces consisting of identical V-shaped nanoparticles, which have different particle number density and arrangement. The resonant second-order nonlinear response of the modified sample was over five times stronger than that of the reference sample. Surprisingly, such a considerable enhancement was achieved while halving the particle number density of the sample array. We believe that these results open new possibilities for the fabrication of more efficient nonlinear metamaterials.

AUTHOR INFORMATION

**Corresponding Author**

*robert.czaplicki@fizyka.umk.pl; martti.kauranen@tut.fi

**Author Contributions**

The manuscript was written through contributions of all authors. All authors have given approval to the final version of the manuscript.

**Funding Sources**

Academy of Finland (265682, 287651, 308596).

# Supporting information for
# Less is more - enhancement of second-harmonic generation from metasurfaces by reduced nanoparticle density

*Robert Czaplicki[†,‡]\*, Antti Kiviniemi[†], Mikko J. Huttunen[†], Xiaorun Zang[†], Timo Stolt[†], Ismo Vartiainen[§], Jérémy Butet[¥], Markku Kuittinen[§], Oliver Martin[¥], and Martti Kauranen[†]\**

[†] Photonics Laboratory, Tampere University of Technology, P. O. Box 692, FI-33101 Tampere, Finland

[‡] Institute of Physics, Faculty of Physics, Astronomy, and Informatics, Nicolaus Copernicus University, Grudziadzka 5/7, 87-100 Torun, Poland

[§] Institute of Photonics, University of Eastern Finland, P. O. Box 111, FI-80101 Joensuu, Finland

[¥] Nanophotonics and Metrology Laboratory (NAM), Swiss Federal Institute of Technology, Lausanne (EPFL), 1015 Lausanne, Switzerland

\*robert.czaplicki@fizyka.umk.pl; martti.kauranen@tut.fi

1. **Extinction spectra**

The linear spectra of the samples were determined by measuring their extinction at normal incidence for *x*- and *y*- polarizations. In order to support the results obtained in the experiments, the linear responses of the samples were simulated using the finite element method (FEM, by COMSOL Multiphysics) (Fig. S1). In the modelling, the substrate and covering layer (20 nm thick above the V-particles) are fully taken into account, as the V-particles are embedded in a medium of refractive index $n \sim 1.45$. In addition, the outlines of the V-particles are extracted from the SEM images to mimic the real geometry of the nanoparticle. The optical properties of the gold nanoparticles are taken from the tabulated Johnson and Christy data [1].

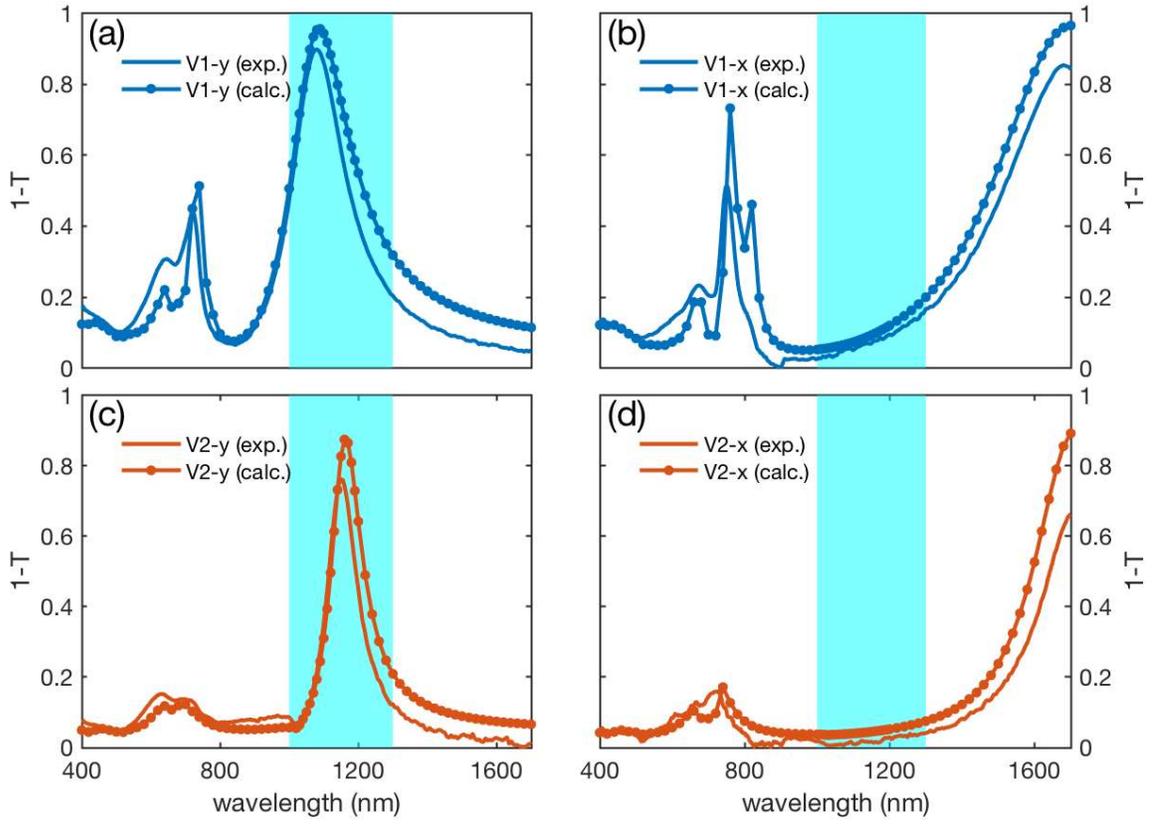

Fig. S5. *y*-, (a), (c) and *x*-polarized, (b), (d) extinction spectra for V1 (blue) (a)-(b) and V2 (red) (c)-(d) samples. The solid lines indicate experimental results and solid-dotted lines are the theoretically calculated spectra. The colored (cyan) area highlights the fundamental wavelength range (1000 – 1300 nm).

## 2. Nonlinear discrete-dipole approximation simulations of all four sample arrays

In addition to samples V1 and V2 presented in the main text, we also studied two additional sample configurations V3 and V4, all schematically shown in Fig. S2.

The two additional samples consisted of identical V-shaped nanoparticles (similar to samples V1 and V2), arranged in arrays with periods ($p_x=p_y=p$) of 707 nm and 1000 nm for V3 and V4, respectively. The measured extinction spectra of all four samples for *x*- and *y*-polarized excitation light are shown in Fig. S3.

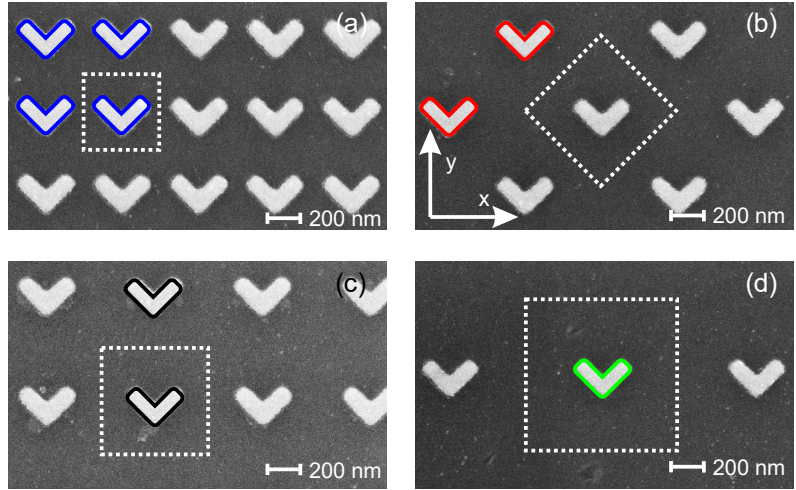

Fig. S6. The scanning electron microscopy images of sample V1 (a) sample V2 (b), sample V3 (c) and sample V4 (d). Nanoparticles outlined by blue (a), red (b), black (c) and green (d) highlight the number of particles inside a 1×1 µm$^2$ area. Dotted-line squares indicate the unit cells of the arrays [500 × 500 nm$^2$ (a), 707 × 707 nm$^2$ (b) and (c) and 1×1 µm$^2$ (d)]. The coordinate system used for all samples is shown as an inset in (b).

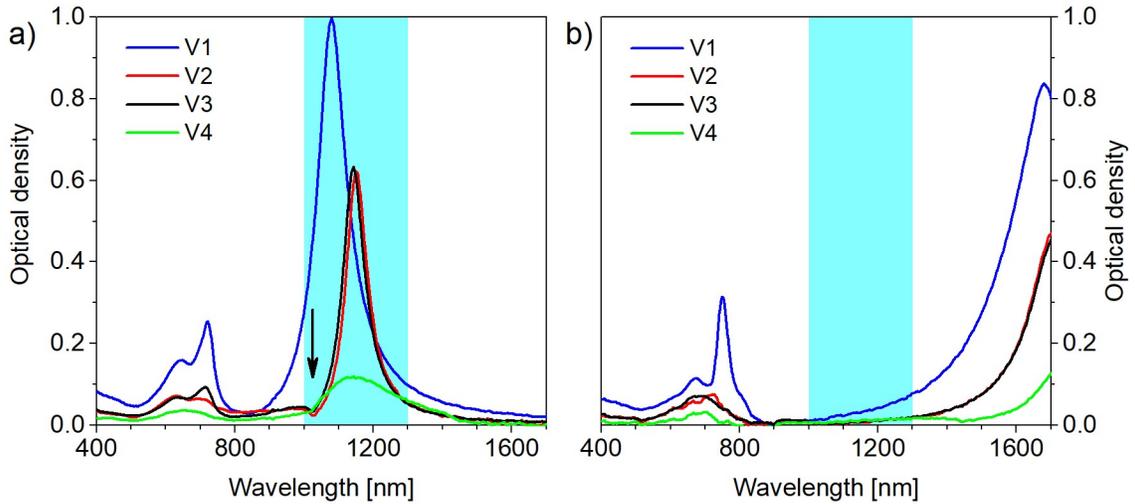

Fig. S7. Measured extinction spectra for samples V1 (blue), V2 (red), V3 (black) and V4 (green) when using incident light polarized along (a) *y*- and (b) *x*-direction. The cyan area highlights the fundamental wavelength range (1000 – 1300 nm) used in the SHG experiments.

In order to verify that the experimental SHG results were due to favorable lattice interactions, we performed numerical SHG simulations using an approach based on nonlinear discrete-dipole approximation (DDA) [2], which is briefly explained also below. In short, a nanoparticle array consisting of *N* particles is assumed to be illuminated by a normally incident linearly polarized (along y-direction) plane wave $\mathbf{E}_{inc}(\omega)$ oscillating at the fundamental frequency $\omega$. This incident field induces a dipole moment $\mathbf{p}(\omega)$ in each of the nanoparticles, which consequently scatters light to the adjacent particles thus affecting their dipole moments. This scattering effect can be taken into account by solving the following system of *3N* linear equations

$$\mathbf{E}_{inc,j}(\omega) = \sum_{k=1}^{N} \mathbf{A}_{jk}(\omega)\mathbf{p}_k(\omega), \qquad (1)$$

where $\mathbf{A}_{jk}(\omega)$ describes the interaction between the $j^{th}$ and the $k^{th}$ nanoparticle and can be written as [3]

$$\mathbf{A}_{jk}(\omega) = \frac{\exp(ikr_{jk})}{\varepsilon_0 r_{jk}}\left[k^2\left(\hat{\mathbf{r}}_{jk}\hat{\mathbf{r}}_{jk} - \mathbf{1}_3\right) - \frac{1-ikr_{jk}}{r_{jk}^2}\left(3\hat{\mathbf{r}}_{jk}\hat{\mathbf{r}}_{jk} - \mathbf{1}_3\right)\right], \quad j \neq k \qquad (2a)$$

$$\mathbf{A}_{jj}(\omega) = \alpha_j^{-1}(\omega), \qquad \text{otherwise} \qquad (2b)$$

where $k = \frac{n\omega}{c}$ is the wavenumber, $n$ is the refractive index of the surrounding medium, $c$ is the speed of light in vacuum, $\varepsilon_0$ is the vacuum permittivity, $r_{jk}$ is the distance between the dipoles (i.e. particles) and $\hat{\mathbf{r}}_{jk}$ is the unit vector pointing from $\mathbf{r}_j$ to $\mathbf{r}_k$. For the case of small nanoparticles, the polarizability $\alpha_j$ of the $j^{th}$ particle can be written as [4]

$$\alpha_j(\omega) = \frac{A_0}{(\omega_{res} - \omega) + i\gamma}, \qquad (3)$$

where $A_0 = 0.32$ cm$^3$ s$^{-1}$ is a constant depending on the scattering cross-section of the particle [5], $\omega_{res} = \frac{2\pi c}{\lambda_{res}} = 1.70 \times 10^{15}$ Hz ($\lambda_{res} = 1115$ nm) is the center frequency of the localized surface plasmon resonance (LSPR) and $\gamma = 1.4 \times 10^{14}$ s$^{-1}$ is the half-width of the LSPR. These numerical values for the parameters were found by fitting simulated extinction spectra to the measured extinction spectra. The simulated arrays were assumed to consist of 90 × 90 lattice sites. Using Eqs. (1) to (3) the self-consistent dipole moment distribution $\mathbf{p}(\omega)$ can now be solved, after which the local field at the fundamental frequency $\mathbf{E}_{\text{loc}}(\omega)$ can be calculated using the relation

$$\mathbf{E}_{\text{loc}}(\omega) = \varepsilon_0^{-1} \boldsymbol{\alpha}^{-1}(\omega) \mathbf{p}(\omega). \tag{4}$$

Once $\mathbf{E}_{\text{loc}}(\omega)$ has been solved, it can be used to drive the SHG process occurring at the $j^{\text{th}}$ particle given by equation

$$\mathbf{p}_{\text{exc},j}(2\omega) = \varepsilon_0 \boldsymbol{\beta}_j(2\omega;\omega,\omega) : \mathbf{E}_{\text{loc},j}(\omega) \mathbf{E}_{\text{loc},j}(\omega), \tag{5}$$

where $\boldsymbol{\beta}_j(2\omega;\omega,\omega)$ is the hyperpolarizability tensor of the $j^{\text{th}}$ particle. We now see that the associated dipole moments $\mathbf{p}_{\text{exc}}(2\omega)$ oscillate at the SHG frequency $2\omega$. However, $\mathbf{p}_{\text{exc}}(2\omega)$ is not yet self-consistent, since we do not take into account the fact that each $\mathbf{p}_{\text{exc},j}(2\omega)$ is also affected by the second-harmonic light scattered from the other particles in the array. In order to find the self-consistent $\mathbf{p}(2\omega)$, we can write and solve another system of *3N* linear equations, given as

$$\mathbf{E}_{\text{exc},j}(2\omega) = \sum_{k=1}^{N} \mathbf{A}_{jk}(2\omega) \mathbf{p}_k(2\omega), \tag{6}$$

where $\mathbf{A}_{jk}(2\omega)$ describes the interaction between the $j^{\text{th}}$ and the $k^{\text{th}}$ nanoparticle at the SHG frequency and the term

$$\mathbf{E}_{\text{exc},j}(2\omega) = \varepsilon_0^{-1} \boldsymbol{\alpha}_j^{-1}(\omega) \mathbf{p}_k(2\omega), \tag{7}$$

has the expected quadratic dependence on the incident fundamental field [see Eq. (5)].

After the self-consistent $\mathbf{p}(2\omega)$ has been solved from Eq. (6), it is straightforward to calculate the quantity of interest, such as the scattering cross-section or the scattered field. In our case, we are interested in calculating the amount of SHG light scattered in the forward direction (along the optical axis or *z*-direction). In our measurement configuration, this is well approximated by calculating the square modulus of the maxima of the emitted SHG field:

$$I_{SHG}(2\omega) \propto |\mathbf{E}_{exc,max}(2\omega)|^2. \tag{8}$$

The expected SHG signal as a function of incident wavelength can now be calculated using Eq. (8). The comparison between the experimental and the calculated results for the four studied sample arrays is shown in Fig. S4. As mentioned in the main text, we believe that the main discrepancy between the measured and simulated curves is due to the reduction of nanoparticles of finite size to point dipoles and from experimental imperfections that are not fully taken into account. In addition, a small feature near (1130 nm) due to favorable diffraction orders [($\pm$1,0) and (0,$\pm$1)] is visible in the simulated SHG spectrum for the sample V2. This was attributed to the fact, that we considered the case with perfectly homogeneous environment.

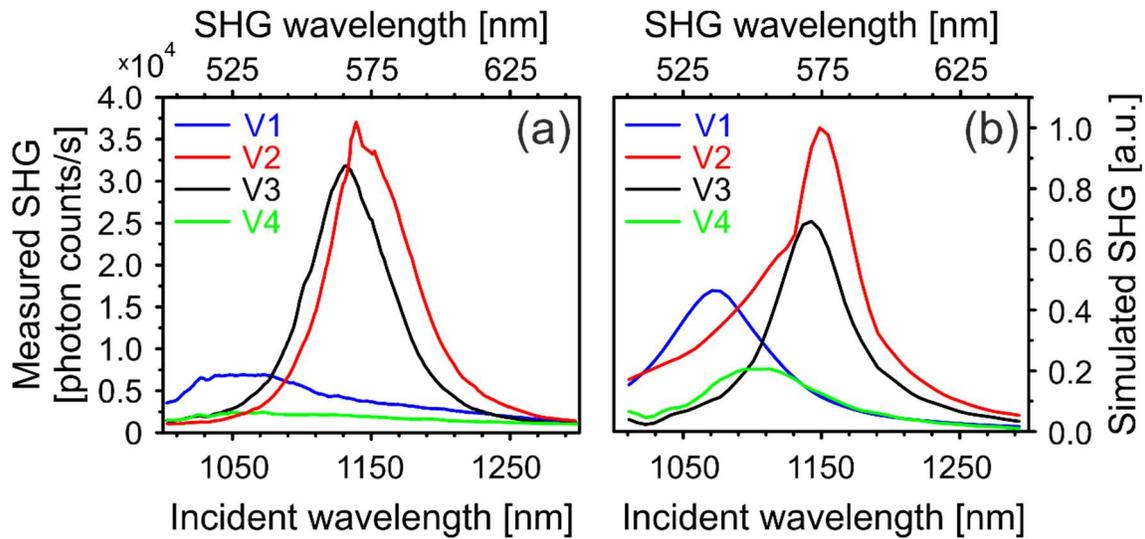

Fig. S8. (a) Experimental and (b) simulated wavelength dependent emitted SHG signal for studied samples V1 (blue), V2 (red), V3 (black) and V4 (green).

### 3. Surface integral equation method

In order to support and verify the performed DDA simulations, we carried out further numerical simulations using surface integral equation method (SIE) to calculate the nonlinear responses of the studied samples. As shown previously [6,7], this is a very powerful tool to calculate SHG response from metamaterials.

We calculated reflectance and the field enhancement at the extremity of the V-arm for the studied samples (Fig. S5). The results show slight redshift in comparison to our experimental results (Fig. 2 in the main text). Nevertheless, we observe clear field enhancement for the sample V2 in comparison with sample V1 [Fig. S5(b)], resulting from the loss modification.

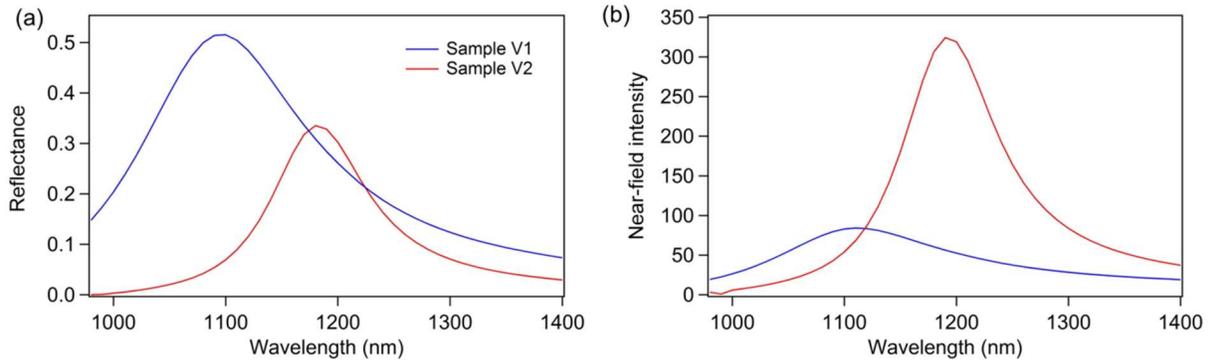

Fig. S5. The reflectance (a) and near-field intensity (b) for samples V1 (blue) and V2 (red).

The modelling of SHG response for the nanoparticles of the exact geometry was not possible due to the proximity of the SH wavelength to the periodicity, which induces the contributions of higher radiative channels. To overcome this difficulty, we reduced the size of the particles. All the geometrical parameters of particles, except the thickness that was kept as original, 20 nm, were decreased by a factor of four resulting in particles of the arm length of 68.75 nm and arm width of 25 nm. The pitch of the lattice 500 nm was also reduced by a factor of four resulting in a square lattice of 125 nm (V1r) and rotated by 45 degrees square lattice of 177 nm (V2r). The reflectance and SHG intensity of the new samples is shown in Fig. S6.

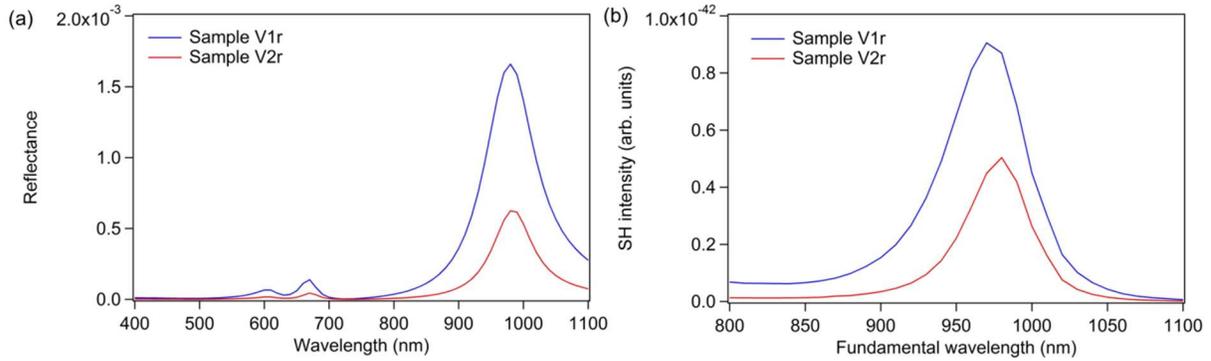

Fig. S9. The theoretically calculated reflectance (a) and second-harmonic intensity (b) of samples with decreased parameters V1r and V2r.

The results from Fig. S6(b) show that the reduction in the number density of particles should result in the lower SHG. These calculations, however, were performed for the samples with different parameters as the ones discussed in the main manuscript. In that case, the resonances for both samples are far away from the possible Rayleigh anomalies (RAs) of the array. In this case it is impossible for the LSPRs to couple to the RAs and form surface-lattice resonances and thus such a result also confirms the importance of lattice interactions on the strong enhancement observed in the main text of the manuscript.

4. **FEM modelling of SHG**

In addition to SIE, we have used FEM to study SHG. The radiated *y*-polarized SHG light was calculated in the forward direction (Fig. S7). As in the case of previous results, also these calculations show an enhancement in the case of sample V2.

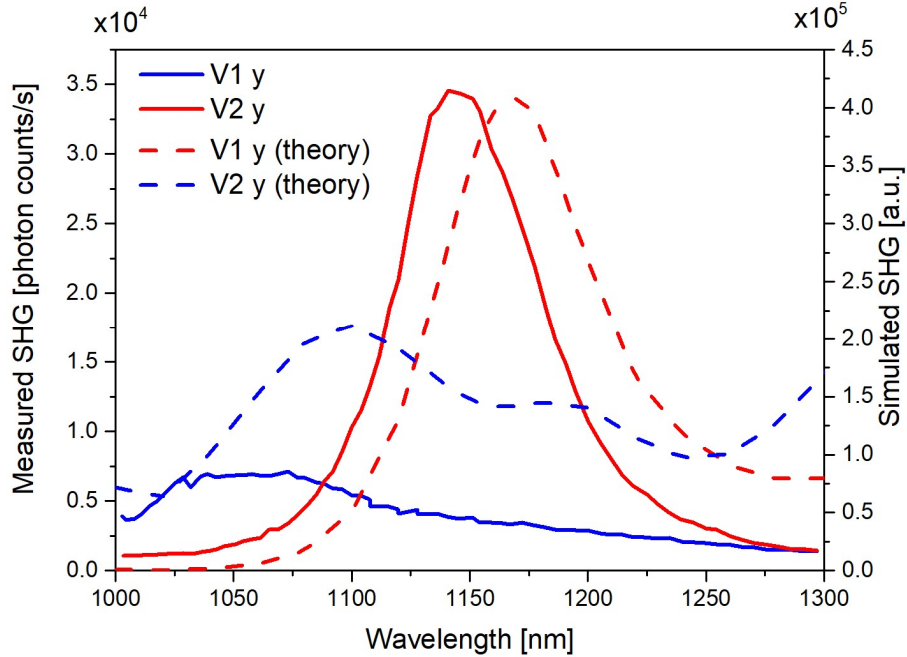

Fig. S10. Experimental (solid lines) and theoretically calculated (dashed lines) SHG emission from samples V1 (blue) and V2 (red). The calculated results show *y*-polarized SHG signal in the forward direction.